\documentclass[twocolumn,aps,showkeys,amsmath,amssymb,epsf]{revtex4}
\usepackage{graphicx}
\usepackage{dcolumn}
\usepackage{color}

\usepackage{bm}
\def\l{\bar{L}}
\def\lt{L_T}
\def\rc{R}
\def\ps{P_s}
\def\lp{L_T-P_s}
\def\ie{i.e.}
\def\eg{e. g.~}
\def\etl{$et ~al.$~}

\def\eqn{\end{equation}\noindent}
\def\eqnr{\end{eqnarray}\noindent}
\def\beqr{\begin{eqnarray}}
 \def\beq{\begin{equation}}

\begin{document}
\title{Understanding the Alternate Bearing Phenomenon: Resource Budget Model}
\author{Awadhesh Prasad$^{1,2}$\footnote{Corresponding author: Tel.:  +91 11 2766 2752, Fax: +91 11 2766 7061,\\
Email:awadhesh@physics.du.ac.in (A. Prasad)\\ 
AP is visiting TUAT under JSPS Fellowship.} 
and Kenshi Sakai$^1$\footnote{Email:ken@cc.tuat.ac.jp (K. Sakai)}}
\affiliation{ 
$^1$Tokyo University of Agriculture and Technology, Tokyo 183-85-9, Japan\\
$^2$Department  of Physics  and Astrophysics, University of Delhi, Delhi 110007, India
}
\begin{abstract} 
We consider here the resource budget model of plant energy resources, which characterizes  the ecological
alternate bearing  phenomenon in fruit crops, in which high and low yields occur in alternate years.
The resource budget model is a tent-type map, which we  study  in detail.
An infinite number of chaotic bands are observed in this map, which are separated by periodic unstable
fixed points. These  $m$ bands chaotic attractors   become $m/2$ bands when the period-$m$
unstable fixed points simultaneously collide with the chaotic bands.
The distance between two sets of coexisting chaotic bands that are separated by a period-$1$  unstable
fixed point is discussed. We explore the effects of varying a range of parameters of the model. 
The presented results explain the characteristic behavior of the alternate bearing estimated from
the real field data.  
Effect of noise are also explored.  The significance of these results to ecological perspectives of
 the alternate bearing phenomenon are highlighted.  
\end{abstract}
\keywords{Resource budget model, Tent map, Band merging crisis, Alternate bearing}
\maketitle

{\bf 
Major plants which produce large seed crops usually show alternate bearing, \ie, 
a heavy yield year is followed by extremely light  ones, and vice versa. This causes the cascading
effect  throughout the ecosystem, and may cause the serious health problem for human 
beings. Therefore, it is very important to understand this natural phenomenon. 
In this paper we attempt to understand some of the  dynamical complexities of this phenomenon.
We consider here a simplest mathematical model which captures many of its characteristic behaviors.
}

\section{Introduction}
Many crop plants of major significance that produce seeded fruit display alternate bearing,
 in which a year of heavy crop yield, known as a mast year, is followed by a year
 of extremely light yield and vice versa.  This causes a cascading
effect  throughout the ecosystem, including leading to serious health problems for human 
beings \cite{koenig1,koenig2,kelly,monse}.  The alternate bearing phenomenon is quite common, and has been  
observed in wild plants in forests as well as domesticated plants.  Understanding the dynamics
 of this natural phenomenon is a difficult task due to the  complexities
 involved in the natural systems, and the limitations of  experimental design and verification. 
In this paper, we attempt to illuminate some of the  dynamical complexities of this phenomenon.

Various attempts have been made to
determine the characteristics of the alternate bearing phenomenon, including, but not limited to,
 discussions in Refs. \cite{koenig1,koenig2,kelly,monse,isagi,ken,satake1,satake2,satake3,ken2,rees,crone,lyles}. 
 However Isagi \etl \cite{isagi}
were the first  to attempt to study the characteristics of the phenomenon using a simple mathematical
model which captures some of its key behaviors.
However, even after this proposition, except for a few attempts 
\cite{isagi,ken,satake1,satake2,satake3,ken2,rees,crone,lyles},
this system has not been studied in detail. In this work, we
endeavor to study this system in detail and correlate the mathematical model with the ecological behavior.
Isagi \etl \cite{isagi} proposed the following model based on the energy resource present in a plant:
a constant amount of photosynthate is produced every year
in individual plant. This photosynthate is used for growth and maintenance of the plant.
{ The remaining 
photosynthate ($P_s$) is stored within the plant body. The accumulated photosynthate stored in the plant 
is expressed as $I$. For a particular year, if the accumulated photosynthate ($I+P_s$) exceed a
 certain threshold $L_T$,
then the remaining amount $I+P_s-L_T$ is used for flowering, with the cost of flowering expressed as $C_f$ 
\ie~$C_f= I+P_s-L_T$.
These flowers are pollinated and bear fruits, the cost of which is designated as $C_a$. Usually, under 
relatively stable conditions, the  fruiting cost is proportional to the cost of flowering,
\ie, $C_a=RC_f$ where $R$ is  a proportionality constant.
 After flowering, the leftover accumulated photosynthate is $L_T$. Once fruiting is over, this is reduced to 
 $I^{\prime} = L_T-C_a$, \ie, $I^{\prime}=-RI+L_T(1+R)-P_sR$.
However, in non-mast years the accumulated energy becomes $I^{\prime}=I+P_s$. This phenomenon, 
considering $I=I_n$ and $I^{\prime}=I_{n+1}$, is modeled \cite{isagi}
 as follows:}

\begin{eqnarray}
\nonumber
I_{n+1}&=&-R I_n+\l    \mbox{~~~~~} I_n>L_T-P_s\\
&=&I_n+P_s  \mbox{~~~~~~~~~} I_n\le L_T-P_s
\label{eq:model}
\end{eqnarray}
where $n=1,2,3....$ represent the years and $\l=L_T(1+R)-P_sR$. 
Here $R$ is  considered as a parameter for this model.
Since this model is based on energy (photosynthate) stored in a  plant it is termed a
 resource budget model \cite{isagi}. 
Here, we explore the characteristic behavior of this model, considering it significance with 
reference to previous discussions \cite{isagi,ken,satake1,satake2,satake3,ken2,rees,crone,lyles}. 
In this work we explain the reason of existence of chaotic bands and their merging. We found that
the $m$ bands chaotic attractors   become $m/2$ bands when the period-$m$
unstable fixed points simultaneously collide with the chaotic bands.

In this model   $L_T$ and $P_s$ are generally considered to be constants for mature trees.
However, the branches of a natural tree may sometimes be cut, or suffer some destruction, which 
implies that there must be some variation in $L_T$. In addition, due to 
weather fluctuations or other sources of external disturbances, the yearly photosynthate production, 
and therefore the quantity $\ps$, also does not 
remain constant. Here, we examine the importance of variations in $L_T$ and $\ps$
to explore their effects on the characteristic behavior of the system, and
correlation with the real-world phenomenon of  alternate bearing. 
{Results show that the ratio of cost of fruiting  to the cost
of flowering, which occurs in multi-bands regime, does not get changed due to
the yearly variations in  photosynthesis (when yearly photosynthesis exceeds certain critical value).}
 However the initial distance between the chaotic bands does depend on photosynthesis.

\begin{figure}
\scalebox{.8}{\includegraphics{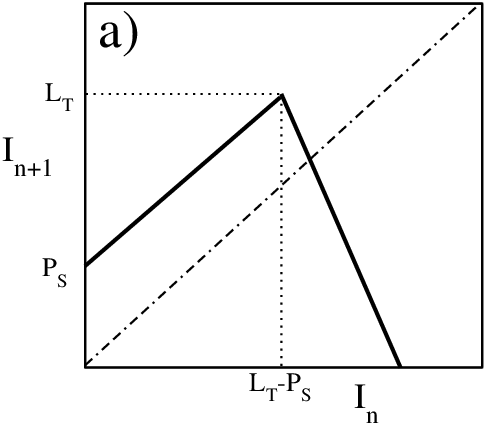}}
\scalebox{.4}{\includegraphics{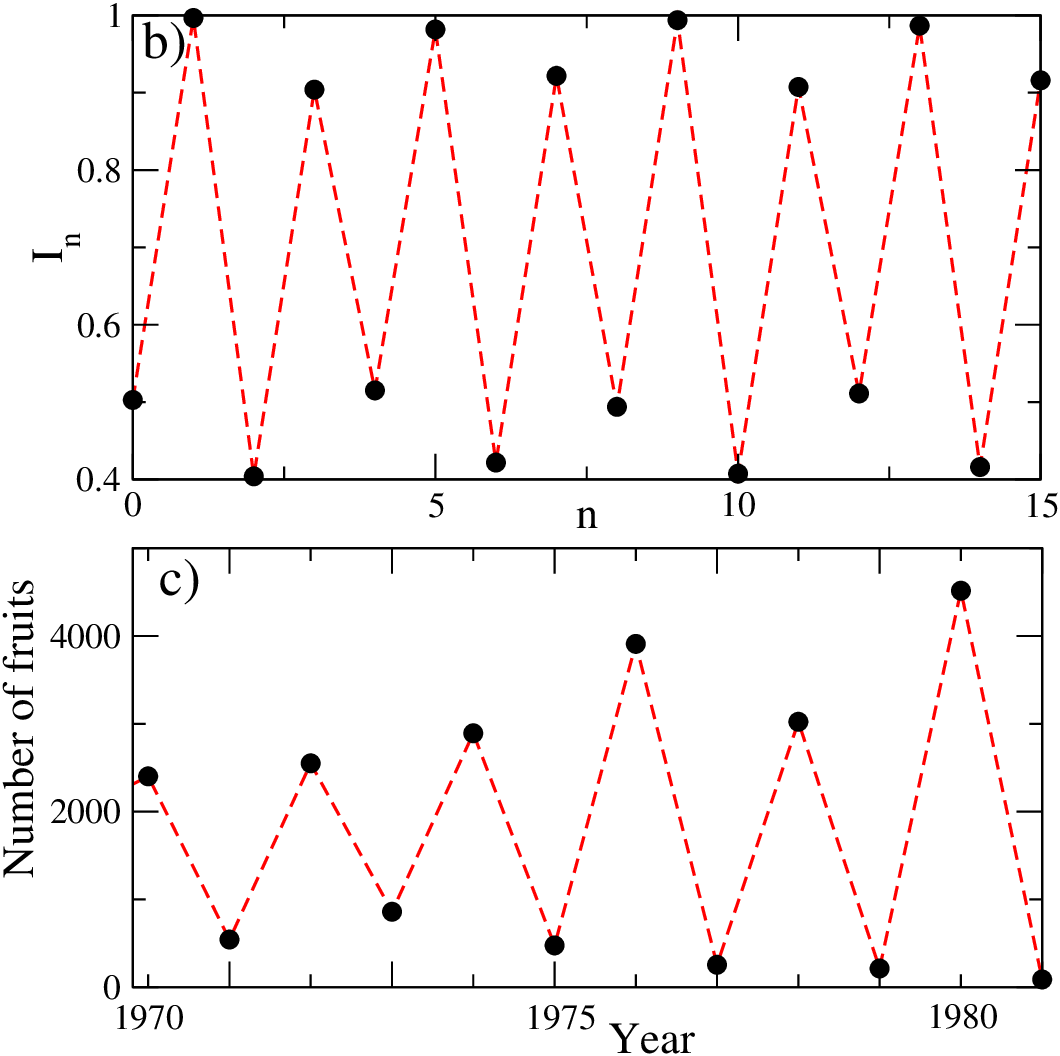}}
\caption{(a) Schematic diagram of the  map of the resource budget model of Eq. (\ref{eq:model}), 
showing the relationship between the stored photosynthate $I_n$ in consecutive years \cite{isagi,ken}.
{ Time-series of (b) model Eq. (\ref{eq:model}) at $R=1.2$ and (c)  number of fruits per year 
of Citrus Unshiu -- see text for details.}} 
\label{fig:model}
\end{figure}

This model looks like as a tent map, as shown schematically in Fig. \ref{fig:model}(a) \cite{ken}.
However, it can be seen that if $I_n$ is below $L_T-P_s$ then it does not depend on parameter $R$ \ie, the
slope is always one.  As discussed below, this independence leads to new results compared to
standard tent maps \cite{tabor,shameri,futter} -- this leads to the creation of separate chaotic 
bands which are useful for understanding the phenomenon of alternate bearing. 
{Note that this is the simplified  model  representing the  phenomenon of alternate bearing. Shown in
Fig. \ref{fig:model}(b) is the time-series generated from this model at $R=1.2$ which clearly indicates the high and low
 fruiting alternatively.  Shown in Fig. \ref{fig:model}(c) is the  number of fruits of an individual Citrus Unshiu tree 
 at the Nebukawa Experimental station in Kanagawa Prefecture over many years. 
This is an example of alternate bearing at individual level whose high and low fruiting is clearly captured 
by the model (Fig. \ref{fig:model}(b)). }

\section{Results and discussion}

\subsection{Generation of Chaos}

\begin{figure}
\scalebox{.6}{\includegraphics{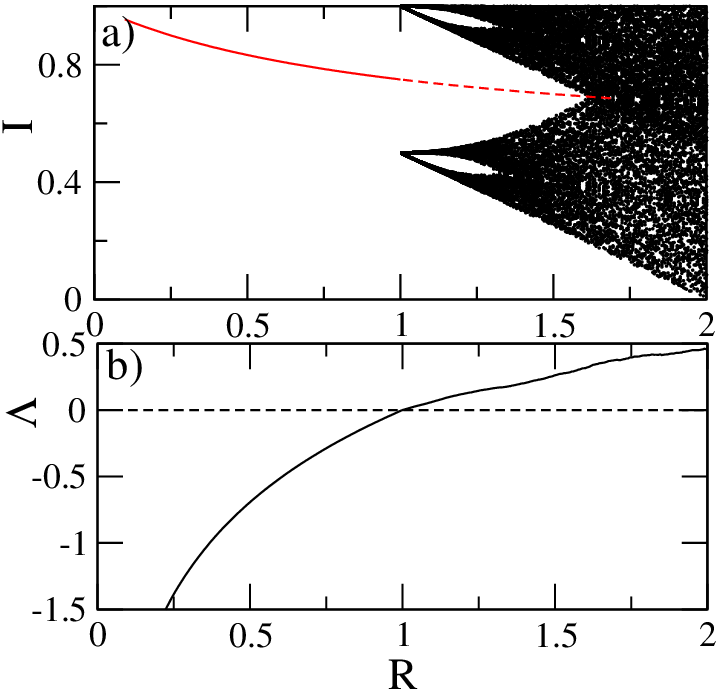}}
\caption{ a) The bifurcation diagram, $I_n$,  and b) the Lyapunov exponent, $\Lambda$ \cite{tabor}, 
 as a function of
parameter $\rc$ for fixed values of  $\lt=1$ and $\ps=0.5$ in  Eq. (\ref{eq:model}). The dashed line in (a)
represents the unstable fixed point of period-1 while in (b) it indicates  $\Lambda=0$.  }
\label{fig:bif}
\end{figure}

Fig. \ref{fig:bif}(a) shows the bifurcation of $I_n$ as a function of $R$ in the model,
Eq. (\ref{eq:model}). At low values of $R$ there is only a single stable solution (shown by the red solid line),
giving the fixed point of period-1, $I_*^{11}$. 
 We will denote the fixed points by $I_*^{mj}$, where $j =1,2,...,m$ represents the sequential order of 
the fixed points of period-$m$.  As $R$ increases above 1, the fixed point $I_*^{11}$ becomes
 unstable (shown by the red dashed line).  Simultaneously,   chaos appears both above  and below the 
period-1 unstable fixed point.  The presence of  chaos (irregular behavior in $I_n$) in these 
trajectories is confirmed 
by determining the Lyapunov exponent $\Lambda$ \cite{tabor}, which is positive  for $R>1$ 
(Fig. \ref{fig:bif}(b)). Here, upper and lower chaotic bands correspond to the high and low yield years
respectively.

\begin{figure}
\scalebox{.6}{\includegraphics{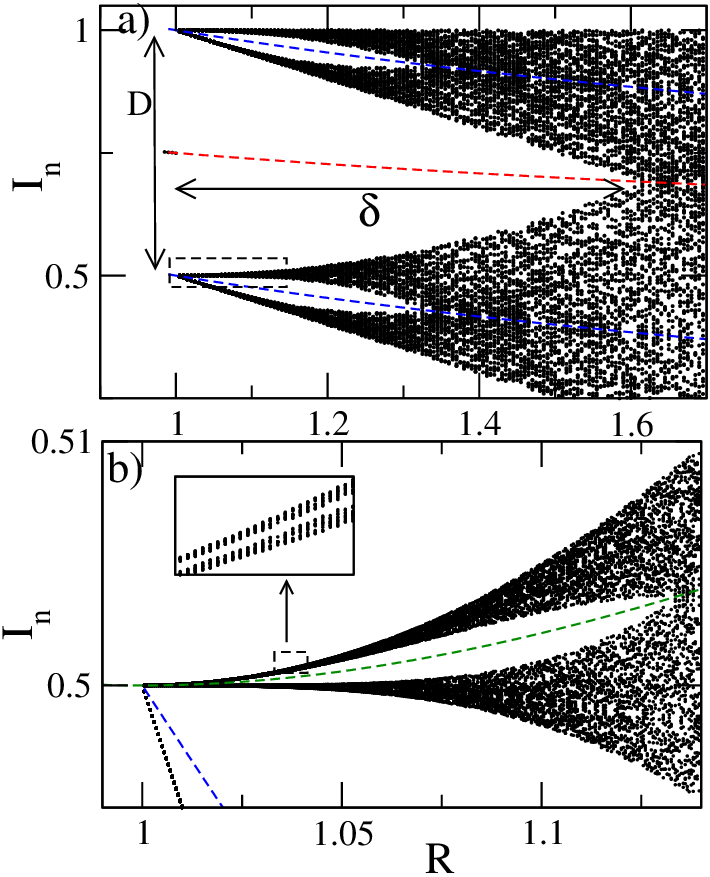}}
\caption{  a) The bifurcation diagram  of Fig. \ref{fig:bif}(a) enlarged in the range $\rc\in [1,1.7]$.  
b) The  expanded view of the dashed box in (a). Inset of (b) corresponds to the small region marked by the   
dashed box.}
\label{fig:small}
\end{figure}

We will now try to understand  the bifurcation, from stable period-$1$ to chaos, in the model in detail.
An expanded view of bifurcation diagram of Fig. \ref{fig:bif}(a)  is shown in Fig. \ref{fig:small}(a), 
over the range $\rc\in[1,1.7]$.
The red dashed line shows the period-$1$ unstable fixed point of the map, Eq. (\ref{eq:model}).
 This fixed point is described by
\begin{eqnarray}
I_*^{11}&=&\frac{\l }{1+R}   >\lp.
\end{eqnarray}
\noindent
The linear stability analysis of Eq. (\ref{eq:model}) around this fixed point shows that the
 eigenvalue $\lambda$ of the stability matrix $M = dI_{n+1}/dI_n$ 
is  $\lambda=R$, since $I_*^{11}>\lp$.
 Therefore, the fixed point is stable while $\rc<1$, and 
unstable for $\rc>1$. This stable and  unstable fixed point is  shown as solid and dashed red lines
respectively in Figs.  \ref{fig:bif}(a) and \ref{fig:small}(b).

At $\rc=1$, two new fixed points of period-2 emerge. These fixed points given by
\begin{eqnarray}
\nonumber
I_*^{21}&=&\frac{\l+P_s }{1+R}    >\lp\\
I_*^{22}&=&\frac{\l-P_sR }{1+R}    \le \lp.
\label{eq:2}
\end{eqnarray}
Linear stability analysis at these periodic fixed points indicates  that the eigenvalue is $\lambda=\rc^2$. 
As such, these period-2 fixed points  are also  unstable for $\rc\ge 1$. The variation of these fixed
 points with $\rc$ is shown as dashed blue lines in Fig. \ref{fig:small}(a). 
These period-2 fixed points pass through two  ``islands"  type of empty space in $I(\rc$).
These islands start near to the respective  fixed points of period-2. Note that there are 
chaotic bands in between the unstable fixed points of period-1 and period-2.

Fig. \ref{fig:small}(b) shows an enlargement of the dashed boxed region in Fig. \ref{fig:small}(b).  Here, 
another, smaller set of empty islands can be seen. The dashed green line shows the trajectory of one of 
the unstable fixed points of
period-4 which pass though these smaller islands. The period-4 fixed points are given by
\begin{eqnarray}
\nonumber
I_*^{41}&=&\frac{\l+\l R^2-\l R+P_s}{1+R^3}    >\lp\\
\nonumber
I_*^{42}&=&\frac{\l+\l R^2-\l R-P_sR}{1+R^3}    >\lp\\
\nonumber
I_*^{43}&=&\frac{\l-\l R+\l R^2+P_sR^2 }{1+R^3}    >\lp\\
I_*^{44}&=&\frac{\l-\l R+\l R^2-P_sR^3 }{1+R^3}    \le \lp.
\label{eq:4}
\end{eqnarray}
\noindent 
Further expansion of the small region marked by a dashed box in Fig. \ref{fig:small}(b) shows another
island, as shown in the inset. A period$-8$ fixed point passes across this island.

\begin{figure}
\scalebox{.5}{\includegraphics{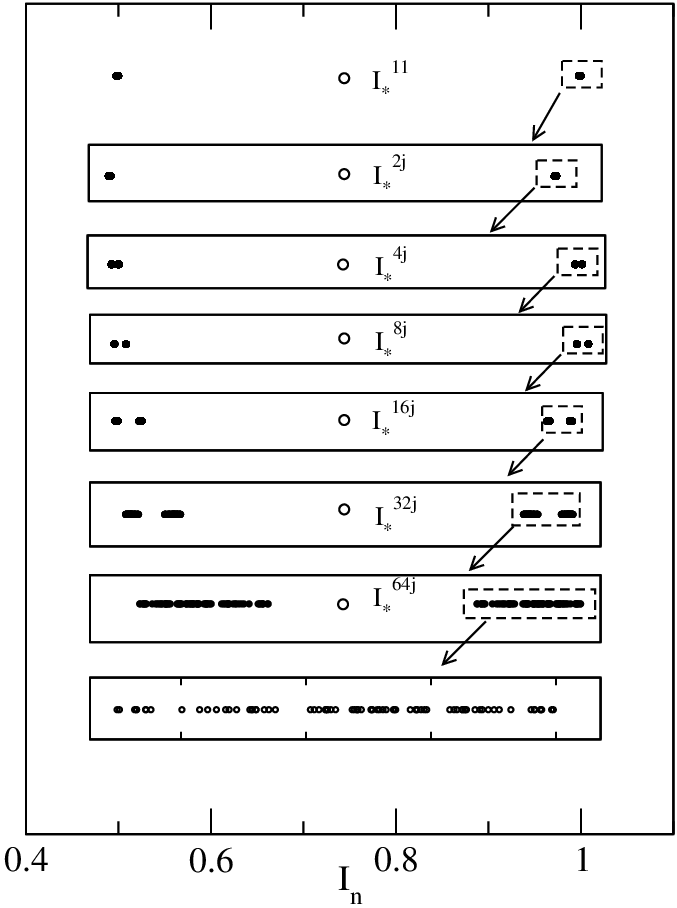}}
\caption{  Positions of phase points, $I_n$ (filled circles) at $R=1.005$, showing bands of chaos. Arrows indicate the expanded view of 
smaller regions (dashed boxes). The positions of one of each of the unstable fixed points, $I_*^{2mj}$, of 
period-$2m$ are shown schematically with open circles.  }
\label{fig:pp}
\end{figure}

In general,  there are an infinite number of islands, which contain  period-$2m$ fixed
points, created near to $\rc=1$. The eigenvalues of these period-$2m$ fixed points are
$\lambda\in[\rc,\rc^{2m}]$, which are  always positive. The exact values depend on the relative values 
of the fixed points with respect to $\lp$, as indicated in 
Eq. (\ref{eq:4}).  Hence,  these fixed points are unstable. This suggests that
an infinite number of chaotic bands are created just after $\rc=1$, separated by
one unstable fixed point of period$-2m$ and period$-2(m+1)$. These bands appear in  very small regions
which are difficult to detect. Shown in Fig. \ref{fig:pp} are the zoomed regions of these bands at $R=1.005$
where we can detect up to period-$64$ fixed points, \ie, $128$-chaotic bands.

\subsection{Separation, $D$, of Chaotic Bands}

To understand the complete dynamics of the system of Eq. (\ref{eq:model}) it
is necessary to determine the effect of  variation of $\ps$. 
All the infinite bands of  chaotic attractors  are  created just after  $\rc=1$.
The bands are separated by period-$2m$ unstable fixed points. 
The two unstable period-2 fixed points lie above and below the unstable period-1 fixed point, separated near to $R=1$ by a 
distance $D$, as shown in Fig. \ref{fig:small}(a).

 $D$ depends only on the
distance between the two unstable fixed points of period$-2$.
Expanding $\l$ in terms of $\rc$, expressions for the two period-$2$ fixed points, Eq. (\ref{eq:2}),
become
\begin{eqnarray}
I_*^{21}&=&\lt    >\lp\\
I_*^{22}&=&\lt-\ps   \le\lp.
\end{eqnarray}
\noindent
This suggests that upper fixed point, $I_*^{21}=\lt$, is independent of $\ps$, while  the lower
one, $I_*^{22}=\lt-\ps $, varies linearly with $\ps$. Therefore, the distance between these two fixed points  is
$D=|I_*^{21}-I_*^{22}|=|\ps|$, \ie, the starting separation of the chaotic bands which appear  above and below the
unstable fixed is dependent on $P_s$ only. Therefore, similar bifurcation diagrams, 
 as those given in Figs. \ref{fig:bif} and \ref{fig:small}, are observed for different values $\ps$ (figures are 
shown here).  Note that the upper fixed point, $I_*^{21}=\lt$ is independent of $\ps$; hence,
irrespective of the values of $\ps$, the fixed point remains at $\lt$.
However, the fixed point $I_*^{22}=\lt-\ps$, which depends on  $\ps$, moves up to $\lt$ as $\ps$ decreases.
 
Ecologically, this dependency of $D$ on $P_s$ suggests  that, if there is high value of 
photosynthate in a particular year (for example, with more sunshine, or less snow fall or clouds) 
then there will be a
huge difference in seed output in the following  years. This means that, as per the above discussion,
the distance $D$, and hence the separation of the upper and  lower chaotic bands will be large.
This simple model thus explains the important ecological phenomenon of alternate bearing, explaining 
origin and variation in magnitude of the effect of the alternate bearing.  This also predicts  that if there is a very high
value of yearly photosynthesis, then one can expect alternate bearing to be observed in the next year. 
{Note that if yearly photosynthesis is less in a plant then there will be small fluctuation in  
$I$ around $L_T$, and hence the the magnitude of variations in yields across the years will also be  less. }

\begin{figure}
\vskip1.cm
\scalebox{0.6}{\includegraphics{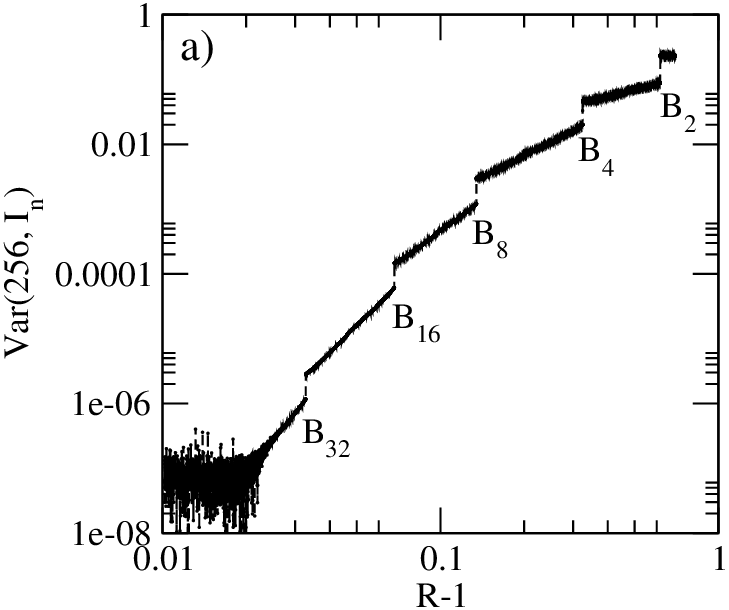}}
\vskip.5cm
\scalebox{0.6}{\includegraphics{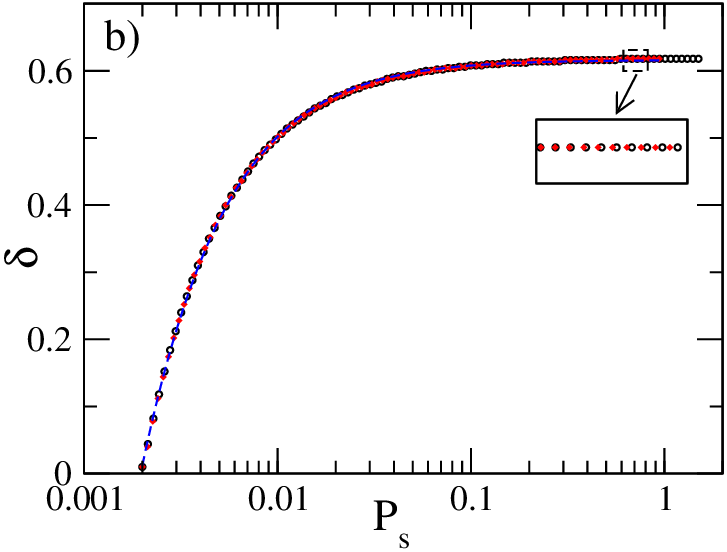}}
\caption{ a) Variance in   $I_n$ after $256$ iterations as a function of $R$.
Positions of $B_{2m}$ indicate the $2m-$bands merging crisis. 
b) Width of  $\delta$ regime as a function of $\ps$.
The black circles and  red squares, correspond to the
 $\lt=1$  and $\lt=1.5$ respectively.  The dashed line is fit to the data by the function Eq. (\ref{eq:fit}).  }
\label{fig:d}
\end{figure}

\subsection{Regime of Multi-Band Chaotic Attractors, $\delta$}

A closer look at  Fig. \ref{fig:small}(a) near to $\rc \sim 1.6$ shows  that
two bands chaos are merged into a single band  when the period-$1$ unstable fixed point collides with the
two bands of   chaotic attractors. This collision  is termed a band-merging crisis \cite{grebogi}.
One method to detect such a crisis is to determine the variance in $I_{n}$ after $2m$-iterations,
\eg within the regime of two bands, the trajectory visits each band alternately, while in the single
merged band the trajectory may move anywhere. Therefore, when a crisis occurs there is large jump in
 the variance of $I_n$.
This is shown in Fig. \ref{fig:d}(a), where a clear jump can be seen at $R\sim 1.6$, when two bands merge.
Similarly, at $\rc\sim 1.3$, four bands of chaotic attractors become two bands, 
as the upper and lower period-2 unstable fixed points each collide with the attractor bands. Also,  
sixteen chaotic bands become 8 bands at the point at which the period-8 unstable fixed points collide separately at lower values
of $\rc$. These band merging crises are shown in Fig. \ref{fig:d}(a). 
 This figure is generated by finding the variance in $I_n$ after $256$ iterations (for detection 
of crises up to the merging of $128$ bands). Note that, due to the infinitely small width of these 
bands (as shown in Fig. \ref{fig:pp}), the higher order merging crisis that it is possible to 
detect numerically is the merging  of the 32 bands containing the period-32 fixed points. 
This shows that, near to $\rc=1$, the $m$ bands chaotic attractors   become $m/2$ bands when the period-$m$
 fixed points simultaneously collide with the chaotic bands. 
The infinite number of chaotic bands created near to  $\rc=1$
 start to merge on further  increase of $\rc$.  
This process of merging  ends when the period-1 fixed point  collides with the remaining two bands of chaos
near to $\rc \sim 1.6$. This regime, in which collision occurs, 
is labeled as $\delta$ in Fig. \ref{fig:small}(a), \ie~ from $\rc=1$ to the  values of $\rc\sim 1.6$ 
where  the two bands of chaos become one band of chaos.
In order to see  how the width of the multi-band regime  $\delta$ changes
with $\ps$,  $\delta$   is numerically determined for different values of $\ps$, as shown in  Fig.  \ref{fig:d}(b). 
 The black circles and red squares correspond to the two values of $\lt=1$ and $\lt=1.5$ respectively.
In the inset to Fig.  \ref{fig:d}(b), it can be seen that the perfect overlapping of these values shows 
that the variation of $\delta$ with $\ps$ is independent of $\lt$.
A fit to the these curves shows  that it is a nonlinear function of the form
\begin{eqnarray}
\delta&=&a(\rc-b)\exp({c \rc^d})+e,
\label{eq:fit}
\end{eqnarray}
where $a, b, c, d,$ and $e$ are fitting parameters.
This function is shown for  $L_T=1$ in  Fig.  \ref{fig:d}(b) as a  dashed blue line, with fitting parameters 
	$a= 8630.07, b = 8632.49, c= -25.945, d= 0.0532924, $ and $e= 0.615928$  .   
	
One important observation is that at  smaller values of $\ps$ $(<0.1)$ there is drastic change in 
$\delta$.  However, for large values it becomes saturated around $\delta=0.61$, \ie, $\rc=1+\delta=1.61$.
Note that the alternate bearing occurs when the values of $R$, \ie, ratio of cost of fruiting  to the cost 
of flowering, is in these multi-bands regime (for $\delta>0$). Since $\delta$ is independent of $\ps$ (after
the  photosynthate, $\ps>0.1$ which is always expected) implies that $R$ does not get changed due to
the yearly variation in  photosynthate. 
One remarkable observation is that this value $\rc = 1+\delta$ is very close to the value of $\rc=1.34$  
estimated from field data \cite{rees}. Therefore the present analysis explains/supports
this important characteristic of ecological alternate bearing phenomenon. 
The slight deviation in these numerical and experimental values  could be due to the external fluctuation 
as discussed below. {Note that this saturation values of $R$ may be different for different type
 of plants, and  may 
depends  on regions and climates \cite{rees} -- which needs to be probed further.}

\begin{figure}
\scalebox{.6}{\includegraphics{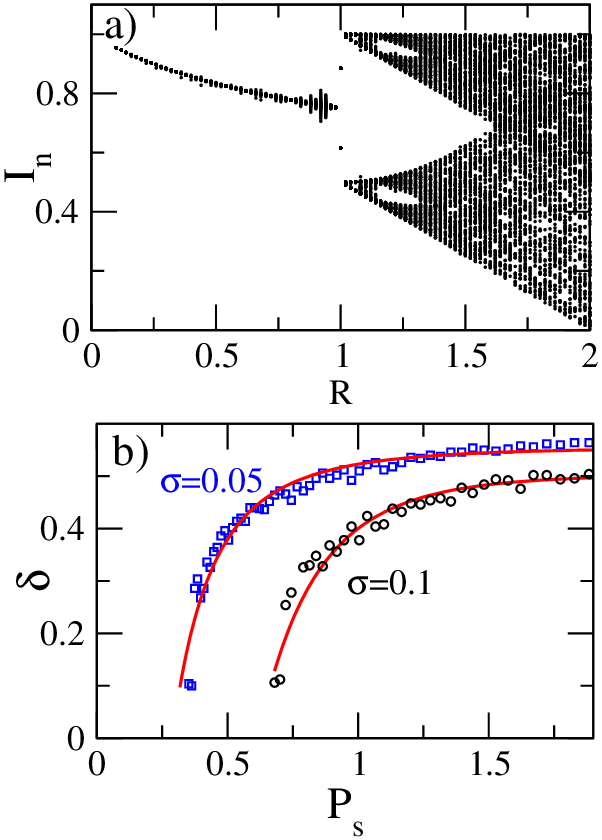}}
\caption{  a) The bifurcation diagram as a function of parameter $\rc$, with noise level $\sigma=0.05$ (5\%).
b) Width   of $\delta$ regime as a function of $\ps$
at noise levels $\sigma=0.05$ (5\%) (blue squares) and $\sigma=0.1$ (10\%) (black circles).
Fittings to the data using function Eq. (\ref{eq:fit}) are shown by solid red lines.  }
\label{fig:noise}
\end{figure}

\subsection{Effects of Noise}

Natural systems are rarely  free from external perturbations. Therefore, in order to fully understand
the dynamics of this system, we must also consider the effect of noise. Here, we consider uniformly distributed 
random noise  in $I_n$, with values between $[0,1]$ and strength $\sigma$. We add this noise  after each
iteration of the map, Eq. (\ref{eq:model}), \ie, $I_n=I_n+\sigma\times[0,1]$. 
Fig. \ref{fig:noise}(a) shows the bifurcation diagram generated with noise of strength $\sigma=0.05$
 ($5\%$ of signal $I$)  for fixed $\lt=1$ and $\ps=0.5$.
 The features remain similar to those without noise, Fig. \ref{fig:bif}(a).
 The  islands and the  distance  $D$  still persist under noise.

To understand the effects of the variations of $P_s$ cause by this noise, two data sets are shown 
in Fig. \ref{fig:noise}(b), with black-circles and blue-squares representing data for 
noise levels $\sigma=0.1$ (10\%) and $\sigma=0.05$ (5\%) respectively. 
 The fits to these datasets, using the function Eq. (\ref{eq:fit}),
 show similar trends to that without noise (Fig. \ref{fig:d}). However, the saturation distance $\delta$ decreases
 for higher strength of noise.
Therefore, the results presented in this work reveals that, under natural conditions in which
fluctuations always exist, the characteristic properties $D$ and $\delta$ 
of alternate bearing phenomenon can persist.

\vskip.1cm    

\section{Conclusions}

Piecewise-smooth dynamical systems, which model  many natural phenomena, are well
studied and found to be important to understand the systems. Piecewise-linear maps have been  also well 
studied (see recent paper \cite{futter} and references therein). Here,  we have studied the
resource budget model which captures the phenomenon of alternate bearing in plants. 
However the map we considered here, 
Eq. (\ref{eq:model}), is different to  standard tent maps \cite{futter,shameri}. In this model, 
one side ($I_n \le \lp$) has a constant coefficient linear equation, while in all reported works
the slopes of both sides change. Therefore this work adds a new class of map to tent-type maps,
 and we were able to study the variations of $D$ and $\delta$.
From a mathematical  point of view, it is important to study this new class of tent map in detail.
This system shows rich bifurcation behavior, where infinite islands containing period-$2m$
unstable orbits are observed. The islands are destroyed due to band merging crises when the
unstable fixed points collide with chaotic bands.

We found that the distance  $D$ only depends on  $P_s$ and independent of $L_T$.
This suggests that if there is a high value of photosynthate in a year then there will be a
huge difference in seed output in the next  year. This variation is independent of the
threshold value $L_T$ of the individual tree.
Therefore, this simple model demonstrates the characteristic properties of an   ecological 
 phenomenon, explaining the magnitude and variation of alternate bearing. 

We observe that only for   smaller values of $\ps$ there is drastic change in
$\delta$.  However, for large values $\delta$ is saturated at approximately $\rc=1.61$, which is very close
 to estimations from real data \cite{rees}.  This shows  that variations in the ratio of cost of 
fruiting to cost of flowering does not dependent on the amount of photosynthate.

These  results, the independence  of $L_T$ on  $D$ and the saturation of regime of multi-bands,
 may be useful for ecological perspectives and hence its open up to new challenges  for further
analysis and verification  in experimental situations. These analysis may also be extended to other types
of important systems \cite{satake1}. It will be also interesting to explore these analysis in
 coupled systems where masting (synchronized production) occurs \cite{isagi,satake1,satake2,rees,satoh}.

\noindent
{\bf Acknowledgment:}
AP acknowledges support from JSPS Invitation Fellowship, Japan and
thanks the TUAT, Fuchu Campus, Tokyo for warm  hospitality.

\end{document}